\documentclass[twocolumn]{revtex4-1}
\usepackage{graphics}
\usepackage{amsmath}
\usepackage{amssymb}
\usepackage{amsfonts}

\begin{document}

\title{A cascade of $e^-e^+$ pair production by a photon with subsequent annihilation to a single photon in a strong magnetic field}

\author{M. M. Diachenko} 
\email{dyachenko.mikhail@mail.ru}

\author{O. P. Novak}
\email{novak-o-p@ukr.net}

\author{R. I. Kholodov}
\email{kholodovroman@yahoo.com}

\affiliation{The Institute of Applied Physics of National Academy of Sciences of Ukraine, 
58, Petropavlivska Street, 40000, Sumy, Ukraine}


\begin{abstract}
The process of electron-positron pair production by a photon with subsequent annihilation to a single photon in a strong magnetic field has been studied. The general amplitude has been calculated and the process rate have been found in the low Landau levels approximation (resonant and nonresonant cases). The comparison of resonant and nonresonant cases shows a significant excess of the resonant rate. The polarization of the final photon in a strong magnetic field has also been found. It has been shown that polarizations of the initial and the final photons are independent except for the case of normal linear polarization of the initial photon.

\vspace{1ex}
\noindent
\textit{2016 Laser Phys.~\textbf{26} 066001;} \textbf{doi:}10.1088/1054-660X/26/6/066001
\end{abstract}

\maketitle

\section{Introduction}

The process of photon propagation in a strong external electromagnetic field remains topical, despite an extensive literature on this subject \cite{Batalin1971}-\cite{Shabad1986}.

The first theory of this process has been developed in Refs.~\cite{Batalin1971}-\cite{Perez1979} using polariztion operator method, and the refractive index of polarized vacuum has been found.
In external electromagnetic field, vacuum becomes an optically active medium, and the effect similar to light birefringence can be observed.
In the works \cite{Shabad2004}-\cite{Hattori2013} considerable attention was paid to vacuum polarization and the singularity of the polarization operator in a constant magnetic field. 
Also, the influence of vacuum polarization on the propagation and absorption of electromagnetic waves near pulsars was considered. 
 
It should be emphasized that in the resonant case the intermediate electron-positron pair still can annihilate to the final photon instead of producing a real electron-positron pair. However, the divergence of the polarization operator is usually associated only with one-photon pair creation. The cascade process of pair production with subsequent annihilation requires more detailed study.
The similar process of production of an electron-positron pair and successive annihilation has been recently considered also in a plane-wave field in \cite{Meuren2015}.
Also, there is another resonance in the problem, which is the result of the formation of a bound state between the created electron and positron \cite{Shabad1985, Shabad1986}.

It should be noted, that the resonant divergences are typical for two vertex QED processes in external electromagnetic field. In this case, the intermediate particle goes to the mass shell \cite{Voroshilo2005}-\cite{Novak2015}.

In this work the process of electron-positron pair production by a photon with subsequent annihilation to a single photon in a magnetic field is studied using diagram technique. 
Magnetic field is considered to be comparable with the critical Schwinger one, ${H_c=m^2c^3/e\hbar=4.41\cdot 10^{13}}$~G.
Such fields are believed to exist at neutron stars \cite{Harding06}-\cite{Harding02}.
Subcritical magnetic field can be generated in heavy ion collisions with the impact parameter comparable to the Compton wavelength of an electron \cite{Fomin_arXiv}. 
The process rate has been found in the Low Landau Levels (LLL-) approximation and the polarization degree of the final photon has been considered in resonant and nonresonant cases.
In should be noted, that the process rate contains the divergence (Dirac delta-function), since there is a single particle in the final state.
Nevertheless, it is still possible to consider photon polarization because the delta function cancels in the expression of polarization degree.

\section{The process amplitude}
Let us consider the propagation of an arbitrary polarized photon in a uniform magnetic field.
The Landau gauge is chosen of electromagnetic 4-potential, which is  $(0;0,xH,0)$.

\begin{figure}
  \resizebox{\columnwidth}{!}{\includegraphics{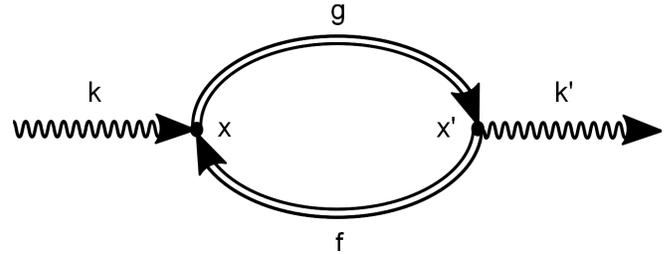}}
  \caption{The Feynman diagram of the process of photon propagation in a magnetic field with production and annihilation of a virtual electron-positron pair}
  \label{diagramma}
\end{figure}

The Feynman diagram of the process is shown in fig.\ref{diagramma}. The wavy line is the photon wave function and the internal double lines are the Green's functions of the virtual electron and the positron in a magnetic field.

The wave function of a photon has a standard form (we will use natural units, where  $\hbar=c=1$, throughout)~\cite{LandauIV}:
\begin{equation}
\label{A}
  A_{\mu} = \sqrt{\frac{2\pi}{\omega V}}e_\mu e^{-ikx},
\end{equation}
where $V$ is the normalization volume, $e_\mu=(0,\vec{e})$ is the polarisation 4-vector of the photon:
\begin{equation}
\label{e}
  \vec e = \left(
  \begin{array}{c}
    \cos\phi \cos\theta \cos\alpha - e^{i\beta} \sin\phi \sin\alpha   \\
    \sin\phi \cos\theta \cos\alpha + e^{i\beta} \cos\phi \sin\alpha   \\
    - \sin\theta \cos\alpha
  \end{array}
  \right).
\end{equation}
Here, $\phi$ and $\theta$ are the azimuthal and polar angles,  $\alpha$ and $\beta$ are the parameters that determine a photon polarization.

The Green's function of an electron in a magnetic field has the following form \cite{Fomin2000}-\cite{Novak2015}
\begin{equation}
\label{G}
  G(x,x') = \frac{-m\sqrt{h}}{(2\pi)^3} 
  \int d^3g e^{-i\Phi} \sum\limits_{n}   \frac{G_H(x,x')}{g_0^2 - E_n^2},
\end{equation}
\begin{eqnarray}
\label{GH}
  G_H(x,x') = (\gamma P + m) \left[\tau U_n U'_n+\tau^* U_{n-1} U'_{n-1}\right]+ \nonumber\\ 
  +im\sqrt{2nh}\gamma^1 \left[\tau U_{n-1} U'_{n}-\tau^* U_n U'_{n-1}\right] ,
\end{eqnarray}
where $h$ is magnetic field strength in the units of the quantum critical magnetic field of $H_c$, $\gamma$ are the Dirac gamma matrices, $U_n$ is the Hermite function, $\rho(x)=m\sqrt{h}x+g_y/m \sqrt{h}$ is the argument of $U_n$ and the primed functions in the formula (\ref{GH}) depend on $x'$,
\begin{eqnarray}
  \Phi = g_0(t-t') - g_y (y-y') - g_z (z-z'), \\
  \tau = \frac{1}{2} (1 + i \gamma_2\gamma_1), \\
  P = (g_0, 0, 0, g_z).
\end{eqnarray}

According to standard rules of quantum electrodynamics \cite{LandauIV}, the amplitude of the process of photon propagation in vacuum in the presence of a magnetic field can be written as
\begin{eqnarray}
\label{Sfi1}
  S_{fi} = e^2 \int d^4x d^4x' 
  \mbox{Tr} [\left( A'^{*}_{2} \gamma\right)G(x',x)\left(A_1\gamma\right) \nonumber \\
  \times G(x,x')],
\end{eqnarray}
where the primed wave function of the final photon depends on $x'$.

Taking into account the expressions for the Green's functions (\ref {G}) and the photon wave functions (\ref {A}) and carrying out corresponding mathematical transformations, it is easy to obtain the following expression for the general amplitude of the process,
\begin{eqnarray}
\label{Sfig}
 S_{fi} = -\frac{4\pi e^2\delta^3(k-k')}{V\omega} \nonumber \\ \times
  \sum_{n_g,n_f=0}^{\infty}\int dg_0 dg_z dg_y \frac{e^{i\Phi'}\sum_{j=1}^6 B_j}{(g_0^2-E_g^2)(f_0^2-E_f^2)} ,
\end{eqnarray}
where $B_j$ have the form:
\begin{eqnarray}
\nonumber
{B_1=e_1e_2(J_{n_f,n_g}^2+J_{n_f-1,n_g-1}^2)(m^2+P'f)},
\end{eqnarray}
\begin{eqnarray}
\nonumber
B_2=(T_1^+T_2^-J_{n_f-1,n_g}^2+T_1^-T_2^+J_{n_f,n_g-1}^2)(m^2+Pf),
\end{eqnarray}
\begin{eqnarray}
\nonumber
B_3=-\sqrt{2n_fh}mg_z(J_{n_f,n_g}J_{n_f-1,n_g}(T_1^+e_2+e_1T_2^-)\\
+J_{n_f,n_g-1}J_{n_f-1,n_g-1}(e_1T_2^++T_1^-e_2)),\nonumber
\end{eqnarray}
\begin{eqnarray}
\nonumber
B_4=\sqrt{2n_gh}m(J_{n_f,n_g}J_{n_f,n_g-1}(e_1T_2^++T_1^-e_2)\\
+J_{n_f-1,n_g-1}J_{n_f-1,n_g}(T_1^+e_2+e_1T_2^-))f_z,\nonumber
\end{eqnarray}
\begin{eqnarray}
\nonumber
B_5=2hm^2\sqrt{n_gn_f}J_{n_f,n_g-1}J_{n_f-1,n_g}(T_1^-T_2^--T_1^+T_2^+),
\end{eqnarray}
\begin{eqnarray}
\nonumber
{B_{6}=4e_1e_2hm^2\sqrt{n_gn_f}J_{n_f,n_g}J_{n_f-1,n_g-1}}.
\end{eqnarray}
Here, 
\begin{eqnarray}
\nonumber
{\Phi'=\frac{k'_x-k_x}{hm^2}g_y-\frac{k_y(k'_x-k_x)}{2hm^2}+(n_g-n_f)(\phi'-\phi)},
\end{eqnarray}
$n_{g,f}$ are the Landau levels of the intermediate particles, $e_{1,2}$ are the parallel to the field components of the polarization vectors of the photons, $J_{n_f,n_g}$ is the known functions appearing in quantum electrodynamics in a magnetic field \cite{Fomin2000}-\cite{Fomin07}.
The following notation is introduced:
\begin{eqnarray}
  P' = (g_0, 0, 0, -g_z),\\
  T_1^{\pm}=\cos\theta \cos\alpha\pm ie^{i\beta}\sin\alpha,\\
  T_2^{\pm}=\cos\theta ' \cos\alpha '\pm ie^{-i\beta '}\sin\alpha '.  
\end{eqnarray}
Note that the functions $T_{1,2}^{\pm}$ determining the polarization properties of photons coincide with the similar functions in the process of two-photon electron-positron pair production in a magnetic field \cite{Diachenko2015}.

After carrying out the integration (\ref{Sfig}) over $dg_y$, the amplitude of the process has the form
\begin{eqnarray}
\label{Sfigd}
 S_{fi} = -\frac{8\pi^2e^2hm^2\delta^4(k-k')}{V\omega} \nonumber \\ \times
  \sum_{n_g,n_f=0}^{\infty}\int dg_0 dg_z \frac{\sum_{j=1}^6 B_j}{(g_0^2-E_g^2)(f_0^2-E_f^2)} .
\end{eqnarray}

Note that the amplitude contains $\delta^4(k-k')$. In other words, in the considered process both energy and momentum conserve, despite the presence of a magnetic field. It should be emphasized that this is not the case for most QED processes in a magnetic field, and the amplitude normally contains only 3-delta function ($\delta(k_x-k'_x)$ is absent for Landau gauge of the potential).

In the considered process the dispersion law is satisfied for the both initial and final photons, $k^2=0$ and $k'^2=0$. Taking into account the 3-delta functions $\delta^3(k-k')$  it is easy to find that $k_x=\pm k'_x$. The case $k_x = -k _x$ should be rejected as non-physical, since it corresponds to reflection of the photon regardless of the magnetic field strength and the photon frequency.

Hereinafter, without loss of generality, we will use a system of reference in which the parallel to the field photon momentum is absent,
\begin{equation}
k_{z}=0,
\end{equation}
since the Lorentz transformation along $\vec H$ does not change the magnetic field.

It is known that the amplitude of this process has the divergence. Therefore, it is necessary to carry out the procedure of regularization or renormalization. We will use the Bogolubov regularization method  \cite {Bogolubov}.
Namely, the denominator in the expression for the Green's function should be replaced as follows:
\begin{eqnarray}
 {({(g,f)}_0^2-E_{g,f}^2+i\epsilon)}^{-1}\to {({(g,f)}_0^2-E_{g,f}^2+i\epsilon)}^{-1}\nonumber\\
 -{({(g,f)}_0^2-E_{g,f}^2-M^2+m^2+i\epsilon)}^{-1}, 
\end{eqnarray}
where $M$ is the additional mass.   
In the Bogolubov $\alpha$ representation \cite{Bogolubov} (which is similar to the Schwinger's proper-time method \cite{Schwinger}), the integrations in (\ref{Sfig}) can be performed using the relation
\begin{equation}
\frac{1}{{g}_0^2-E_{g}^2+i\epsilon}=\frac{1}{i}\int_0^{\infty}{d\tau e^{i\tau ({g}_0^2-E_{g}^2+i\epsilon)}}. 
\end{equation}

After carrying out the procedure of regularization, the general amplitude of the process is given by the formula
\begin{eqnarray}
\label{Sfig2}
  S_{fi} = i\frac{8\pi^3 e^2 hm^2\delta^4(k-k')}{V\omega} \nonumber \\
   \times \sum_{n_g,n_f=0}^{\infty}  \int_0^1 d\zeta \left(a - \frac{b}{2}ln\left|\frac{m^2\zeta(1-\zeta)}{\omega^2(\zeta-\zeta_1)(\zeta-\zeta_2)}\right|\right. \nonumber \\
   -\left.\frac{m^2}{\omega^2}\frac{c(\zeta)}{(\zeta-\zeta_1)(\zeta-\zeta_2)-i\frac{\epsilon}{\omega^2}}\right) .
\end{eqnarray}
Here, $\epsilon$ is the positive small quantity. 
The following notation is used:
\begin{eqnarray}
a=e_1e_2(J_{n_f,n_g}^2+J_{n_f-1,n_g-1}^2) \nonumber\\
+T_1^-T_2^+(J_{n_f-1,n_g}^2+J_{n_f,n_g-1}^2),
\end{eqnarray}
\begin{eqnarray}
{b=2T_1^-T_2^+(J_{n_f-1,n_g}^2+J_{n_f,n_g-1}^2)},
\end{eqnarray}
\begin{eqnarray}
c(\zeta)=a+4e_1e_2h\sqrt{n_g n_f}J_{n_f,n_g}J_{n_f-1,n_g-1} \nonumber\\
+a(1+2n_f h)+2haN_{-}\zeta,
\end{eqnarray}
\begin{eqnarray}
\zeta_{1,2}=\frac{1}{2}-\frac{m^2}{\omega^2}hN_{-} \nonumber\\
\pm\sqrt{\frac{\omega^2}{m^2}(\frac{\omega^2}{m^2}-4[1+N_+ h])+4h^2 N_-^2},
\end{eqnarray}
$N_{\pm}=n_g\pm n_f$ and $\omega$ is the frequency of the initial photon.

\section{The resonant conditions and the process rate}

Let us consider the resonant case when virtual particles become real ones. 
In this case,  both first-order poles in (\ref{Sfig2}) coincide with each other: 
\begin{equation}
\label{res}
\zeta_1=\zeta_2=\zeta_{res}.
\end{equation}
Equation (\ref{res}) has two solutions, one of which is spurious. Indeed, $\zeta_{res} $ does not belong to the integration interval (\ref{Sfig2}), hence, the singularity in the integrand vanishes. Therefore, we will use the other solution, which has the form
\begin{equation}
\label{wres}
\omega_{res}=m\left(\sqrt{1+2n_g h}+\sqrt{1+2n_f h}\right).
\end{equation}
Thus, the resonant frequency is equal to the sum of the electron and the positron energies with zero longitudinal momenta.
It should be noted that the same condition was obtained in \cite{Shabad1975} when analyzing the polarization operator in the presence of a magnetic field.
 
In the resonant case (\ref{wres}) the amplitude of the process (\ref{Sfig2}) takes on the form
\begin{eqnarray}
  S_{fi}^{res} = -i\frac{8h\pi^4 e^2 m^4}{V\omega^2\Gamma}\delta^4(k-k') \sum_{n_g,n_f=0}^{\infty} c(\zeta_{res}) ,
\end{eqnarray}
where $\Gamma$ is the width of the resonance.

Hereinafter, we will use LLL-approximation,
\begin{equation}
  \label{LLL}
  n_{g,f}\sim 1, \quad 
  hn_{g,f}<<1.
\end{equation}
These conditions are common in subcritical magnetic field, when the field strength approaches the critical value of $H_c$.
In this approximation in the first order in $h$ the resonant photon frequency can be rewritten as:
\begin{equation}
\omega_{res}=m(2+N_+ h).
\end{equation}

In the view of the above expressions, the resonant amplitude in the linear approximation on $h$ has the form
\begin{eqnarray}
\label{Sfi5}
  S_{fi}^{res} = -i\frac{2h\pi^4 e^2 m^2}{V\Gamma}\delta^4(k-k')\nonumber \\ \times
  J^2(e_1 e_2 [2+h(N_+ -2n_g n_f)]+hN_+ T_1^- T_2^+) ,
\end{eqnarray}
where 
\begin{equation}
\label{J}
  {J=\frac{(-1)^{n_f}}{\sqrt{n_g! n_f!}}e^{-\frac{\eta}{2}}\eta^{\frac{N_+}{2}}}.
\end{equation}
Here, $\eta=(k_{x}^2+k_{y}^2)/2hm^2.$

It should be noted that based on the formula for the amplitude (\ref{Sfi5}) the optical theorem can be derived. In other words the rate of single-photon electron-positron pair production is determined by the imaginary part of the amplitude of the studied process. In the case of unpolarized photon, the optical theorem looks like
\begin{equation}
\label{Op}
  W_{pair}=-\frac{2\Gamma}{\sqrt{m\delta \omega}} ImS_{fi}^{res},
\end{equation}
where $W_{pair}$ is the probability of the process of one-photon electron-positron pair creation (see for example \cite{Diachenko2015, Novak09}), $\delta \omega$ is the detuning from the threshold of the one-photon pair creation. It should be emphasized that in the formula (\ref{Op}) $\delta \omega$ goes to zero if the pair is created on fixed Landau levels with zero longitudinal momenta. 

It should be noted that the methods of quantum field theory (QFT) \cite{Larkin59}-\cite{Khelemelya}, in particular the optical theorem, can be applied to the problem of heavy charged particle passing through the magnetized electron gas, which is related to the method of electron cooling.
In this method the emittance of heavy charged particle beam  is reduced  due to collisions with electrons which have a small velocity spread  \cite{Budker78}-\cite{Parkhomchuk00}.
The energy loss of a charged particle in the first Born approximation is determined by the imaginary part of the polarization operator. 
It should be noted that despite the widespread usage of the electron cooling method in accelerating technology, yet there is no complete theoretical description of the difference in the friction force for positive and negative particles, which is experimentally observed  \cite {Dikanskii88}. 
This problem will be important for electron cooling of antiproton beams at the Hight Energy Storage Ring (HESR) \cite{Galnander09, Bazhenov03}. 
The difference in cooling of antiprotons and protons can be described using QFT methods in the second Born approximation.

Let us find the resonance rate of the process of photon propagation in a strong magnetic field. 
According to the well-known QED rules the differential rate is determined as
\begin{equation}
\label{dW}
  dW=|S_{fi}|^2\frac{Vd^3k}{(2\pi)^3}.
\end{equation}
The integration over $d^3 k$ can be carried out using $\delta$-function properties. 
The rate accurate to the first corrections in $h$ can be obtained in the form
\begin{eqnarray}
\label{Wres}
  W_{res}=\frac{\pi h^2 m^4 e^4}{2^5\Gamma^2}\delta(\omega-\omega')J^4
  \left((1+\xi_3)(1+\xi'_3) \right. 
  \nonumber \\ \times
  \left.
  [1+(N_+ -2n_g n_f)h]+(\xi_1 \xi'_1+\xi_2 \xi'_2)N_+ h\right),
\end{eqnarray}
where $\xi_{1,2,3}$ are the Stokes parameters of photons \cite{Tolhoek}:
\begin{eqnarray}
\nonumber
\xi_1=\sin 2\alpha \cos\beta,
\end{eqnarray}
\begin{eqnarray}
\nonumber
\xi_2=\sin 2\alpha \sin\beta,
\end{eqnarray}
\begin{eqnarray}
\nonumber
\xi_3=\cos 2\alpha.
\end{eqnarray}

It follows from (\ref{Wres}) that the rate substantially dependent on the initial photon polarization and vanish in the case of normal linear polarization ($\xi_3=-1$) within the accuracy of the approximation.
Also, the suppression of process rate occurs in the case $\xi=-1$ in other processes in a magnetic field as well (for example, in the process of electron-positron pair by a single photon \cite{Novak2012}. The account of the next correction in $h$ gives a small but non-zero rate,  
\begin{equation}
\label{Wxi}
  W_{res}^{(\xi_3=-1)}=\frac{\pi h^4 m^4 e^4}{2^6\Gamma^2}\delta(\omega-\omega')J^4 N_+^2(1-\xi'_3).
\end{equation}

\begin{figure}
  \resizebox{\columnwidth}{!}{\includegraphics{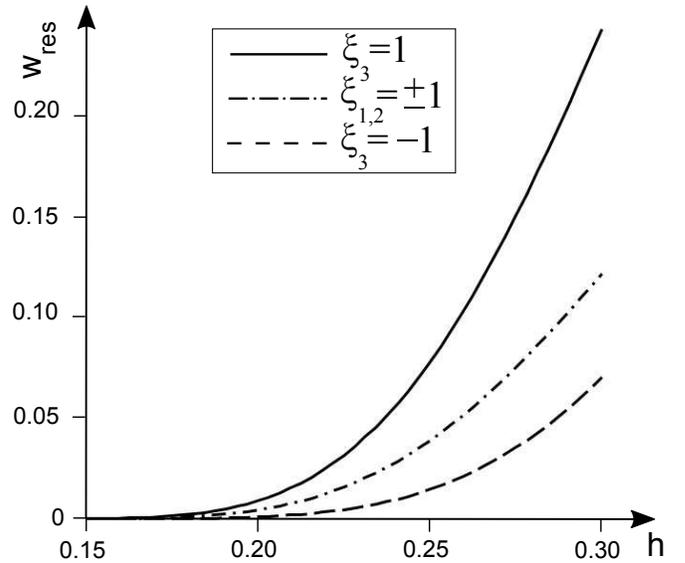}}
  \caption{The dependence of the dimensionless process rate on $h$.}
  \label{Whres}
\end{figure}

As it is known, the two vertex processes have the resonant divergences. In this case, the intermediate particle 
goes to the mass shell and second-order process can be represented by two subsequent first-order processes.
The resonant process of photon propagation in a magnetic field can be represented as a sequence of processes 
of one-photon pair creation and annihilation of pair in a single photon. According to \cite{Novak2012}, in the case when the 
photon has normal linear polarization ($\xi=-1$), the rate of the process of pair production by a photon is 
suppressed. Thus, the rate of the process {(Fig. 1)} is also suppressed when the initial photon has normally 
linear polarized.

It can be concluded from the expressions (\ref{Wres}) and (\ref{Wxi}) that the rate is considerably less in the later case. 
Fig.(\ref{Whres}) shows the dependence of the dimensionless process rate on the magnetic field in the units of the critical strength for various initial photon polarizations. 
It should be noted that the width of resonance has the form \cite{Diachenko2015}
 \begin{equation}
\label{Gamma}
  \Gamma = \frac{2}{3}e^2 m h^2.
\end{equation}
%

\section{Nonresonant case}

Let us consider the process in the nonresonant case when the condition (\ref{res}) is not fulfilled. 
In this case, the initial photon frequency looks like
\begin{equation}
\label{wnotres}
  \omega=\omega_{res}+\kappa m h,
\end{equation}
where $\kappa$ belongs to the interval $(0,1)$.

Taking into account the general amplitude and the condition (\ref{wnotres}) the nonresonant amplitude in LLL-approximation (\ref{LLL}) takes on the form
\begin{eqnarray}
\label{Sfi4}
  S_{fi}^{nonres} = \frac{h\pi^4 e^2 m^2}{V\omega\sqrt{\kappa h}}\delta^4(k-k')J^2 (s_1e_1e_2 \nonumber \\
  +s_2T_1^-T_2^+) ,
\end{eqnarray}
where
\begin{eqnarray}
\nonumber
s_1=1+\frac{24i}{\pi}\sqrt{\kappa h}-h(\frac{13}{8}\kappa+\frac{5}{4}N_+ +n_g n_f),
\end{eqnarray}
\begin{eqnarray}
\nonumber
s_2=\frac{hN_+}{2}.
\end{eqnarray}

Similarly to the previous section, one can find the rate of the nonresonant process:
\begin{eqnarray}
\label{Wnres}
  W_{nonres}=\frac{\pi h m^2 e^4}{2^{11}\kappa}\delta(\omega-\omega')J^4(1+\xi_3)(1+\xi'_3).
\end{eqnarray}

In the same way, when the initial photon has normal linear polarization the nonresonant process rate is
\begin{equation}
\label{Wnresxi}
  W_{nonres}^{(\xi_3=-1)}=\frac{\pi h^3 m^2 e^4}{2^{12}\kappa}\delta(\omega-\omega')J^4 N_+^2(1-\xi'_3).
\end{equation}

The formulas (\ref{Wnres}), (\ref{Wnresxi}) show that the rate in the case of normal linear polarization of the initial photon is much less then in other cases.

According to the expressions (\ref{Wres}), (\ref{Wnres}), the ratio of the resonant and nonresonant rates can be found as
\begin{equation}
\frac{W_{res}}{W_{nonres}}=\frac{144\kappa}{e^4 h^3}.
\end{equation}
It follows from the above expressions that the process rate is substantially greater in the resonant case compared to the nonresonant one.


\section{Polarization of the final photon }

Let us find the polarization of the final photon when the resonance condition (\ref{res}) is fulfilled.
By the definition, the degree of polarization of the final photon can be written as
\begin{equation}
\label{P1}
P = \frac{W(\xi')-W(-\xi')}{W(\xi')+W(-\xi')}.
\end{equation}
After developing Eq. (\ref{P1}) into a power series and keeping terms linear in $h$, the degree of polarization takes on the form
\begin{equation}
\label{P2}
P=\xi'_3+\frac{\xi_1\xi'_1+\xi_2\xi'_2}{1+\xi_3}N_+ h.
\end{equation}

Equality (\ref{P2}) shows that the degree of polarization is almost independent on the polarization of the initial photon.
Taking into account the relation (\ref{P2}), the Stokes parameters of the final photon can be written as 
\begin{equation}
\xi'_3=1,~~\xi'_1=\frac{\xi_1}{1+\xi_3}N_+ h,~~\xi'_2=\frac{\xi_2}{1+\xi_3}N_+ h.
\end{equation}
Thus, the final photon polarization is almost always linear anomalous and $P \approx 1$.

The case ${\xi_3=-1}$ is the only exception. 
According to the expression (\ref{Wxi}), the  polarization of the final photon is  ${\xi'_3=-1}$. Therefore, the photon propagates in a magnetic field without changing the polarization when the initial polarization is normal linear one.
It should be noted that the photon polarization shows similar behavior in the nonresonant case.

Thus, if the photon is polarized as an eigenmode (normal and anomalous mode polarizations are $\xi=-1$ and $\xi=1$, respectively), its polarization does not change after traveling through the magnetic field, in accordance with known results [1-3].


\section{Conclusion}

In the present work the process of photon propagation in a subcritical magnetic field has been considered. 
The photon polarization is assumed to have arbitrary values. 
The obtained expressions of the process rate have been analyzed in the LLL-approximation. 

Considerable attention has been paid to the resonant case (\ref{wres}) when the virtual electron-positron pair goes to the mass shell.
It should be emphasised that the resonant photon propagation with production and annihilation of electron-positron pair is possible when the condition (\ref{wres}) is true.
In the previous works, however, the resonance was associated only to the single-photon production of a real pair.
The expressions for the process rate in the resonant and nonresonant cases substantially depend on the initial photon polarization. 
In the case of normal linear polarization ($\xi_3=-1$) the rate has the minimum value. 

It was found from the consideration of the polarization degree, that the final photon has anomalous linear polarization (${\xi'_3 = 1}$) in most cases.
The exception is the case of normal linear polarization ${\xi_3=-1}$, when the photon propagates without changing of the polarization ($\xi'_3=-1$).
Thus, vacuum in a magnetic field does not show optical activity when the photon polarization $\xi_3=\pm 1$.


\end{document}